\documentclass{article}
\usepackage{amsmath,amssymb}
\usepackage{epsfig} 

\newcommand{\textfrac}[2]{%
  \ensuremath{{}^{#1}\hspace{-.2em}/\hspace{-.1em}_{#2}}}
\newcommand{\ket}[1]{\left| #1 \right\rangle}                              
\begin{document}

\title{Modeling the Singlet State\\ with Local Variables}
\author{Jan-{\AA}ke Larsson\footnote{Electronic address:
    jalar@mai.liu.se}\\
    Matematiska Institutionen,\\
    Link\"opings Universitet,\\ 
    S-581 83 Link\"oping, Sweden.}
\date{December 14, 1999}
\maketitle
\begin{quote}
  A local-variable model yielding the statistics from the singlet
  state is presented for the case of inefficient detectors and/or
  lowered visibility.  It has independent errors and the highest
  efficiency at perfect visibility is $77.80\%$, while the highest
  visibility at perfect detector-efficiency is $63.66\%$. Thus, the
  model cannot be refuted by measurements made to date.
\end{quote}

\section{Introduction}
The singlet state
\begin{equation}
  \ket{\psi}=\tfrac{1}{\sqrt{2}}(\ket{\uparrow\downarrow}
  -\ket{\downarrow\uparrow})
  \label{eq:singlet}
\end{equation}
is the most common choice of quantum state in statements concerning
violations of local realism.  This state is chosen for its special
properties that may be used to test for such a violation using the
Bell inequality \cite{Bell64}

\begin{equation}
  \bigl| E(AB')-E(AC') \bigr| \le 1+E(BC'), \label{eq:bell}
\end{equation}

\begin{figure}[htbp]
  \begin{center}
    \psfig{file=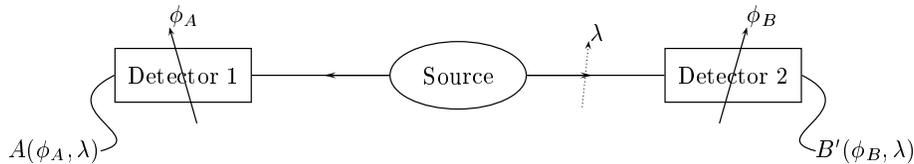,width=\linewidth} 
    \caption{The experiment setup and some
      associated notation.  The source emits a pair of
      spin-\textfrac{1}{2} particles in the singlet state, and
      $\lambda$ denotes the ``hidden variable'' to be used in the
      local hidden-variable model.}
    \label{fig:detectors}
  \end{center} 
\end{figure}

The notation is presented in Fig.~\ref{fig:detectors}, where the
experimental setup is schematically shown.
In ineq.\ (\ref{eq:bell}) the notation is somewhat shortened, so that
$A=A(\phi_A,\lambda)$ and $B=B(\phi_B,\lambda)$ pertains to the first
detector while $B'=B'(\phi_B,\lambda)$ and $C'=C'(\phi_C,\lambda)$
pertains to the second, and the expressions $E(AB')$, $E(AC')$, and
$E(BC')$ are correlations of the results at detector~1 and detector~2
for different detector orientations.  The results are labeled ``spin
up'' ($\uparrow$ or $+1$) and ``spin down'' ($\downarrow$ or $-1$),
and the key property of the singlet state in this context is the total
anticorrelation of the measurement results when the detectors are
equally oriented. In the notation in Fig.~\ref{fig:detectors},
\begin{equation}
  B'(\phi_B,\lambda)=-A(\phi_A,\lambda)\;\text{ if }\;\phi_B=\phi_A,
  \label{eq:anticorr}
\end{equation}
except on a set of zero probability.

If a local realistic (or ``local hidden-variable'') model is possible,
the inequality (\ref{eq:bell}) holds\footnote{A formal statement of the
prerequisites of the inequality is available in e.g.\ 
\cite{Jalar98a}.}, and the conclusion is that if the inequality is
violated, no such model is possible.  Since the derivation of the
statistics of the singlet state is available in many standard
textbooks on quantum mechanics, it will not be repeated here.  The
probabilities are (with $\theta=\phi_B-\phi_A$):
\begin{gather}
  \addtocounter{equation}{1}
  P_{++}^{\psi}=\frac{1-\cos(\theta)}4
  \tag{\theequation a}\label{eq:QM1}\\
  P_{+-}^{\psi}=\frac{1+\cos(\theta)}4
  \tag{\theequation b}\\
  P_{-+}^{\psi}=\frac{1+\cos(\theta)}4
  \tag{\theequation c}\\
  P_{--}^{\psi}=\frac{1-\cos(\theta)}4,
  \tag{\theequation d}\label{eq:QM4}
\end{gather}
and the correlation obtained from this is
\begin{equation}
  E^{\psi}(AB')=P_{++}^{\psi}-P_{+-}^{\psi}
  -P_{-+}^{\psi}+P_{--}^{\psi} =-\cos(\theta).
\end{equation}
Needless to say, the correlation obtained from quantum mechanics
violates the Bell inequality, so there can be no local hidden-variable
model (in the ideal case).

The original spin-\textfrac{1}{2} formulation will be retained in this
paper, whereas present experiments in this context are most often
performed with correlated photons which yield similar statistics for
results labeled ``horizontal polarization'' ($\leftrightarrow$ or
$+1$) and ``vertical polarization'' ($\updownarrow$ or $-1$). Quite a
number of experiments of this kind have been performed (see e.g.\ 
Refs.~\cite{Asp1,Asp2,Asp3,WJSWZ}), and although many experiments to
date are reported to violate the Bell inequality, there are some
problems, most notably detector inefficiency and lowered visibility
(see e.g.\ 
Refs.~\cite{Jalar98a,CHSH,ClauHorn,ClauShim,GargMerm})\footnote{There
  are other experimental issues, but these will not be discussed here.
  See e.g.\ Ref.~\cite{WJSWZ} where the authors enforce the locality
  prerequisite.}.  The efficiency problem concerns the possibility of
missing a detection altogether, while the visibility problem concerns
the possibility of getting an erroneous detection, i.e., a detection
that is converted from $+1$ to $-1$ or vice versa by an error. These
two problems have a slightly different effect on the inequality.

Lowered efficiency does not change the correlation at all, provided
the errors are independent and single events are discarded (events
where both particles are missed are discarded automatically, so to
speak). Therefore this would not seem to be a problem.  Unfortunately,
the inequality itself is influenced, as the original Bell inequality
is simply not valid in an inefficient experiment.  Letting
$\eta$ be the efficiency of the detectors, and $E_{AB'}(AB')$ be the
correlation where the single events (and no-detection events) have
been removed, it is possible to derive the following generalized
inequality \cite{Jalar98a}:
\begin{equation}
  \bigl|E_{AB'}(AB')-E_{AC'}(AC')\bigr|
  \le\frac{4}{\eta}-3+E_{BC'}(BC'),
\end{equation} 
In the form given here, the inequality is valid for constant
efficiency and independent errors.  As is evident from the above
inequality, when $\eta$ decreases, the inequality becomes less
restrictive on the correlations. The inequality is violated by the
singlet state when $\eta>\textfrac{8}{9}\approx 88.89\%$, but using
the Clauser-Horne-Shimony-Holt (CHSH) inequality \cite{CHSH} the bound
is lower \cite{Jalar98a,GargMerm}:
\begin{equation}
  \bigl|E_{AC'}(AC')-E_{AD'}(AD')\bigr|
  +\bigl|E_{BC'}(BC')+E_{BD'}(BD')\bigr| \le\frac{4}{\eta}-2.
  \label{eq:chshn}
\end{equation} 
Here, the bound is $\eta>2(\sqrt{2}-1)\approx 82.83\%$ (actually, this
bound is necessary and sufficient \cite{GargMerm}).

Lowered visibility is also a problem in the Bell inequality, because it
needs full visibility to retain the total anticorrelation in
(\ref{eq:anticorr}).  However, the CHSH inequality does not need
(\ref{eq:anticorr}) and is not changed by lowered visibility, so this
inequality will be used below. The effect is instead that the
amplitude of the correlation function is lowered, and using the
parameter $v$ as a measure of the visibility,
\begin{equation}
  E^{\text{nonideal}}_{AB'}(AB')=-v\cos(\theta),\label{eq:QMnoni}
\end{equation}
which lowers the violation of the CHSH inequality. Since we have
removed the single (and no-detection) events in the correlation, the
efficiency $\eta$ does not appear in the above expression (but the
effect of a lowered $\eta$ is included in the inequality
(\ref{eq:chshn})). 

In the nonideal case, we have the following probabilities for constant
detector efficiency and independent errors:
\begin{gather}
    P(\text{detection of one particle})=\eta,\\
    P(\text{detection of both particles})=\eta^2.
\end{gather}
At either of the detectors, the single-particle probabilities are equal
\begin{equation}
  P_+=P_-=\frac{\eta}{2},
  \label{eq:one}
\end{equation}
and on measurement setup as a whole, the double-particle probabilities
are
\begin{gather}
  \addtocounter{equation}{1}
  \label{eq:QMn1}
  P^{\text{nonideal}}_{++}=\eta^2\frac{1-v\cos(\theta)}4
  \tag{\theequation a}\\
  P^{\text{nonideal}}_{+-}=\eta^2\frac{1+v\cos(\theta)}4
  \tag{\theequation b}\\
  P^{\text{nonideal}}_{-+}=\eta^2\frac{1+v\cos(\theta)}4
  \tag{\theequation c}\\
  P^{\text{nonideal}}_{--}=\eta^2\frac{1-v\cos(\theta)}4
  \tag{\theequation d}.\label{eq:QMn4}
\end{gather}

In the construction below, the case of full visibility will be
investigated as a starting point, leaving the case of lowered
visibility to Section~\ref{sec:vis}. In Section~\ref{sec:straight},
the bounds in $\eta$ and $v$ for the model will be compared to the
bounds in the Bell (or really, the CHSH) inequality. Finally, some
concluding remarks will be made in Section~\ref{sec:concl}.

\section{Local variables: Detector inefficiency}
 
The model is inspired by a simple type of model often seen in the
literature, where the hidden variable $\lambda$ is simply an angular
coordinate (e.g.~\cite{Baggott92,Santos96}), and the measurement
results and the detection errors depend only on the orientation of the
detector with respect to $\lambda$. In the below construction, an
extension is made so that $\lambda$ is the pair $\lambda=(\phi,r)$
consisting of an angular variable $\phi\in[0,2\pi]$, which may be
thought of as the ``spin orientation'', along with another variable
$r\in[0,1]$, the ``detection parameter''.  For each pair, the hidden
variable $\lambda$ has a definite value, and the whole ensemble is
taken to be an even distribution in the rectangle
$(\phi,r)\in[0,2\pi]\times[0,1]$.

\begin{figure}[htbp]
  \begin{center}
    \psfig{file=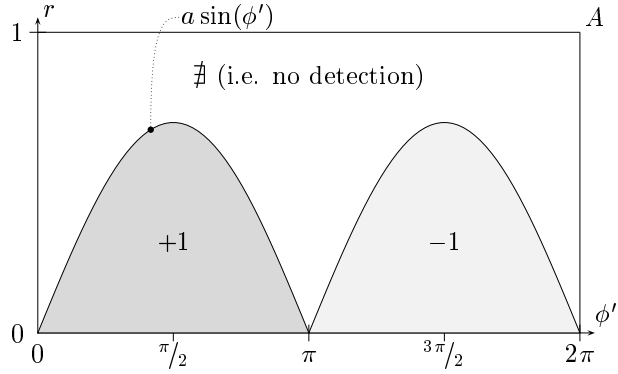} 
    \caption{The detector pattern at detector 1. To get the sinusoidal 
      form of the probabilities, the detector pattern should be
      sinusoidal at least at one detector.} \label{fig:sinus1}
  \end{center} 
\end{figure}

\begin{figure}[htbp]
  \begin{center}
    \psfig{file=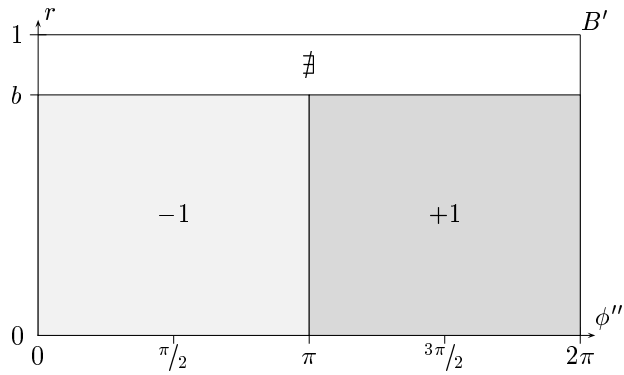} 
    \caption{The pattern at detector 2.} \label{fig:sinus2}
  \end{center} 
\end{figure}

To visualize the model and to simplify the presentation, a series of
figures will be used, where the ``detector patterns'' will show the
experimental results for different values of $\lambda$. The sinusoidal
form of the probabilities is the first aim to fulfill in the model, so
initially the other desired properties will not be present in the
model (see Figs.~\ref{fig:sinus1} and~\ref{fig:sinus2}).

The measurement results (and the detection errors) are given in the
figures, and the procedure to obtain the measurement result is as
follows: at detector 1, the value of $\phi$ is shifted to
$\phi'=\phi-\phi_A$ ($r$ is not changed), and the result is read off
in Fig.~\ref{fig:sinus1}. 

At detector 2 the procedure is similar, but
in this case the shift is $\phi''=\phi-\phi_B$. The result is $+1$
$(-1)$ when $\lambda$ falls in an area marked $+1$ $(-1)$, resp..  If
$\lambda$ should happen to fall in an area marked $\nexists$, a
measurement error (non-detection) occurs and the random variable $A$
($B'$) is not assigned a value (c.f.\ Ref.~\cite{Jalar98a}).

\begin{figure}[htbp]
  \begin{center}
    \psfig{file=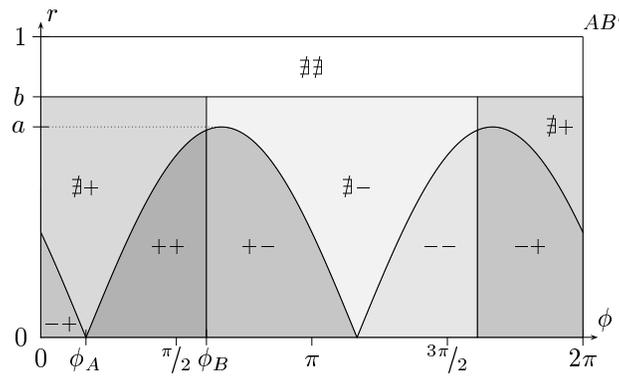} 
    \caption{The detector patterns from Figs.~\ref{fig:sinus1} 
      and~\ref{fig:sinus2} interposed in the $\lambda$-plane.}
    \label{fig:calc}
  \end{center} 
\end{figure} 

\begin{figure}[htbp]
  \begin{center}
    \psfig{file=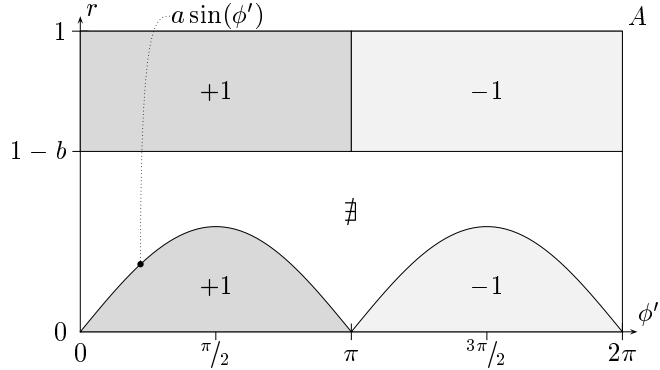} 
    \caption{The detector pattern for detector 1 in the symmetrized 
      model.}
    \label{fig:symm1}
  \end{center} 
\end{figure}

\begin{figure}[htbp]
  \begin{center}
    \psfig{file=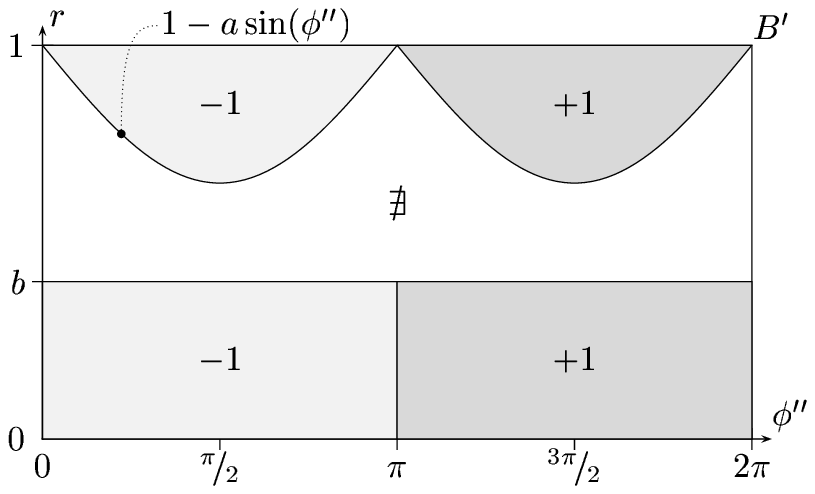} 
    \caption{The detector pattern for detector 2 in the symmetrized 
      model.}
    \label{fig:symm2}
  \end{center} 
\end{figure}

The probabilities for the coincident detections are now possible to
calculate using Fig.~\ref{fig:calc}, where the detector patterns have
been interposed in the $\lambda$-plane with the proper shifts.  E.g.,
the probability of getting $A=B'=+1$ is the size of the area marked
``$++$'' (the size is the area divided by $2\pi$, because the total
probability is one, whereas the total area is $2\pi$, also use
$\theta=\phi_B-\phi_A$):
\begin{equation}
  P_{++}=\frac{1}{2\pi}\int_0^{\theta}a\sin(\phi)d\phi
  =\frac{a(1-\cos(\theta))}{2\pi},\label{eq:prob}
\end{equation}
which has the sinusoidal form we want. Note that the calculation above
is only valid if 
\begin{equation}
  0\le a\le b\le 1, 
\end{equation}
and it is easy to see that the probability would not be sinusoidal if
$a>b$. Thus, the model has the required form of the probabilities and
constant efficiency, but an unwanted property is that the efficiencies
of the detectors are different:
\begin{equation}
  \eta_{B'}= b,
\end{equation}
while 
\begin{equation}
  \eta_A= \frac{2}{2\pi}\int_0^{\pi}a\sin(\phi)d\phi
  =\frac{2a}{\pi}\le\frac{2b}{\pi}<b.
\end{equation}
To resolve this, one way is to lower the efficiency of detector 2 by
simply inserting random detection errors at that detector (which would
yield the model described in \cite{Santos96}), but there is another
way of resolving the problem, by \emph{symmetrizing} the model (see
Figs.~\ref{fig:symm1} and~\ref{fig:symm2}).


In this model, the parameters $a$ and $b$ are subject to the
conditions 
\begin{equation}
  \label{eq:abcomp}
  0\le a\le b\le\textfrac12.
\end{equation}
The efficiency is obviously constant at
\begin{equation}
  \eta= \eta_A=\eta_{B'}= \frac{2a}{\pi}+b.
\end{equation}
The errors should be independent, i.e.,
\begin{equation}
  \eta^2= \frac{4a}{\pi},
\end{equation}
and we have 
\begin{equation}
  \left\{
  \begin{split}
    a&=\frac{\pi\eta^2}{4}\\
    b&=\eta-\frac{\eta^2}{2}.
  \end{split}
  \right.
\end{equation}

\begin{figure}[htbp]
  \begin{center}
    \psfig{file=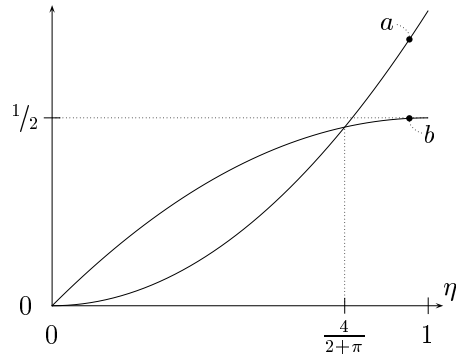} 
    \caption{Comparison of $a$ and $b$ (i.e., $a\le b$).}
    \label{fig:ab}
  \end{center} 
\end{figure} 

By using (\ref{eq:abcomp}) we arrive at the comparison of $a$ and $b$ in
Fig.~\ref{fig:ab}, and thus, the interval at which we may use this
model is
\begin{equation}
  0\le \eta \le \frac{4}{2+\pi} \approx0.7780, \label{eq:eff}
\end{equation}
i.e., the model described above is usable up to $77.80\%$ efficiency.
The probability of ``++'' in the symmetrized model is twice that in
eqn.\ (\ref{eq:prob}) by the symmetrization, which is
\begin{equation}
  P_{++} =2\frac{a(1-\cos(\theta))}{2\pi}
  =\eta^2\frac{1-\cos(\theta)}4,
\end{equation}
as prescribed in (\ref{eq:QMn1}) using $v=1$, and the other
probabilities are checked in the same manner.

\section{Local variables: Visibility} \label{sec:vis}

\begin{figure}[htbp]
  \begin{center}
    \psfig{file=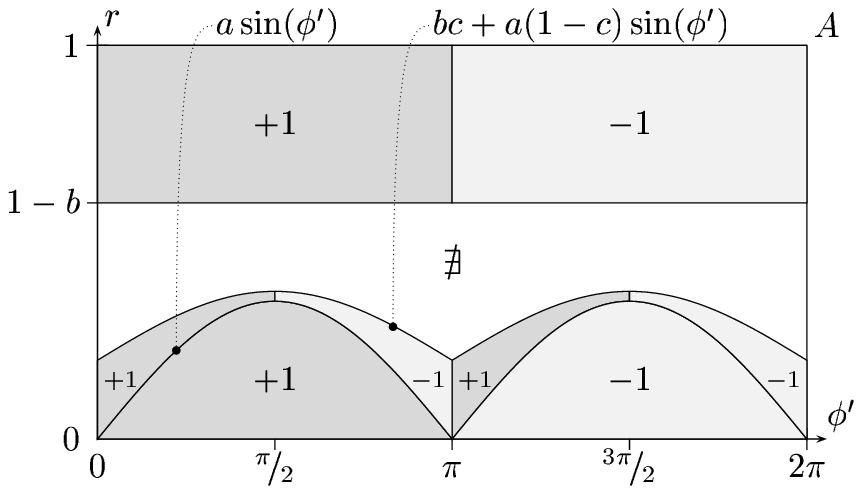} 
    \caption{The detector pattern for detector 1 in the symmetrized 
      model with lowered visibility. The detector pattern for detector
      2 is symmetric in a similar fashion as in the previous section.}
    \label{fig:vis1}
  \end{center} 
\end{figure}

Even though present experiments regularly have a high visibility
(e.g.\ $97\%$ in Ref.~\cite{WJSWZ}) it would be interesting to include
this effect in the model presented in the previous section.  The
visibility is lowered in the model by adding random erroneous
detections in it (see Fig.~\ref{fig:vis1})\footnote{The errors may be
  added as random errors on the whole of the area marked $A\,\nexists$
  in Fig.~\ref{fig:symm1}, but to write the model as a
  ``deterministic'' model and to simplify the calculations, the errors
  are added in the manner as in Fig.~\ref{fig:vis1}. There is no
  effect on the statistical properties of the errors by this choice,
  i.e., the errors are independent and the probability of an error is
  constant.}.

%\begin{figure}[htbp]
%  \begin{center}
%    \psfig{file=modvis2.eps} 
%    \caption{The detector pattern for detector 2 in the symmetrized 
%      model with lowered visibility.}
%    \label{fig:symm2}
%  \end{center} 
%\end{figure} 

Also in this model, the parameters $a$ and $b$ are subject to the
conditions
\begin{equation}
  \label{eq:abcomp2}
  0\le a\le b\le\textfrac12,
\end{equation}
while the parameter $c$ is subject to the condition $0\le c \le 1$ and
is the amount of errors added in the model, so that when $c=0$ we have
full visibility.  The sinusoidal pattern is not changed, in order not
to change the general behavior of the probabilities. The efficiency
is constant at
\begin{equation}
  \eta=b+\bigl(bc+\frac{2a(1-c)}{\pi}\bigr).
\end{equation}
while the independence of the errors yields
\begin{equation}
  \eta^2=2\bigl(bc+\frac{2a(1-c)}{\pi}\bigr).
\end{equation}
The visibility is the amount of ``correct results'' at equal detector
orientations:
\begin{equation}
  v=\frac{4a}{\pi}\Big/\eta^2,
\end{equation}
and after a simple calculation 
\begin{equation}
  \left\{
  \begin{split}
    a&=\frac{\pi v\eta^2}{4}\\
    b&=\eta-\frac{\eta^2}{2}\\
    c&=\frac{\eta(1-v)}{2-\eta(1+v)}.
  \end{split}
  \right.\label{eq:abc}
\end{equation}
The probability of the result ``++'' is obtained by the following
expression, where the limits of the first integral is obtained by the
symmetry of the model
\begin{equation}
  \begin{split}
    P_{++}&=\frac{2}{2\pi}\int_0^{\frac{\pi}2}
    \bigl(bc+a(1-c)\sin(\phi)\bigr)-a\sin(\phi)d\phi\\
    &\qquad+\frac{2}{2\pi}\int_0^{\theta}a\sin(\phi)d\phi\\
%    &=\frac12\Bigl(bc+\frac{2a(1-c)}{\pi}\Bigr)-\frac{a}{\pi}
%    +\Bigl(\frac{a}{\pi}-\frac{a}{\pi}\cos(\theta)\Bigr)\\
    &=\frac12\Bigl(bc+\frac{2a(1-c)}{\pi}\Bigr)
    -\frac{a}{\pi}\cos(\theta)\\
    &=\frac{\eta^2}{4} -\frac{v\eta^2}{4}\cos(\theta),
  \end{split}\label{eq:prob2}
\end{equation}
exactly as in (\ref{eq:QMn1}).
  
\begin{figure}[htbp]
  \begin{center}
    \psfig{file=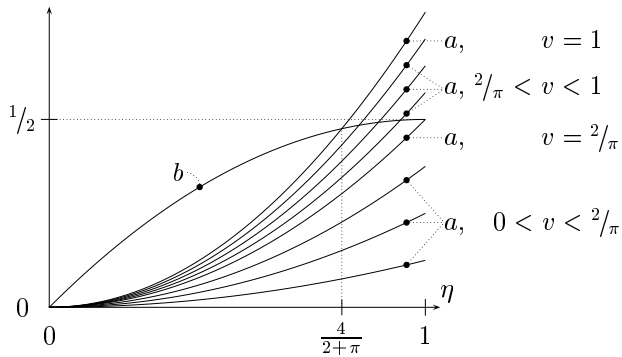} 
    \caption{Comparison of $a$ and $b$ in the model including 
      visibility (i.e., $a \le b$).}
    \label{fig:abv}
  \end{center} 
\end{figure}
Examining the parameters, $b$ is as before and the change in $a$ is an
extra factor $v$, and comparing $a$ and $b$ (see Fig.~\ref{fig:abv})
we can see that if $\eta$ is less than the $77.80\%$ discussed in the
previous section, there is no restriction on the visibility. If $\eta$
is larger than this bound, the interval at which we may use this model
is (use (\ref{eq:abc}) in (\ref{eq:abcomp2}))
\begin{equation}
  v\le \frac{1}{\pi}\bigl(\frac{4}{\eta}-2\bigr).\label{eq:vis}
\end{equation}
This excludes the point $\eta=1$, $v=1$, where $c$ is on the form
\textfrac00, and for all other values of $\eta$ and $v$, $0\le c \le
1$.  One may especially note that if $\eta=1$, the interval is
\begin{equation}
  v\le \frac{2}{\pi}.
\end{equation}

\section{The role of the Bell inequality} \label{sec:straight}

In the Bell inequality (or really the CHSH inequality), the $82.83\%$
is obtained by choosing detector orientations so that the violation of
the inequality is maximized:
\begin{equation}
  \phi_A=0,\;\phi_B=\textfrac{\pi}2,\;\phi_C=\textfrac{\pi}4, 
  \text{ and }\phi_D=\textfrac{3\pi}4.
\end{equation}
With these choices of angles in the CHSH inequality, and using
$E^{\text{nonideal}}_{AB'}$ from (\ref{eq:QMnoni}), we obtain
$2\sqrt{2}v\le4\eta^{-1}-2,$ or
\begin{equation}
  v\le \frac1{2\sqrt{2}}\bigl(\frac4{\eta}-2\bigr),
\end{equation}

\begin{figure}[htbp]
  \begin{center}
    \psfig{file=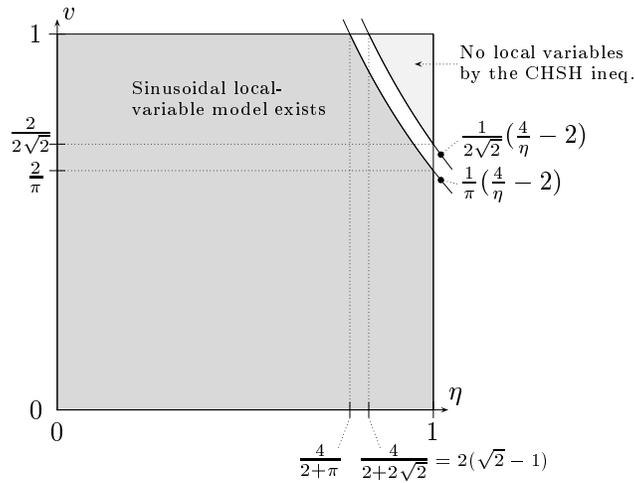} 
    \caption{The region where the model is usable as 
      compared to the region where no local hidden-variable model is
      possible.}
    \label{fig:nv1}
  \end{center} 
\end{figure}
 
\begin{figure}[htbp]
  \begin{center}
    \psfig{file=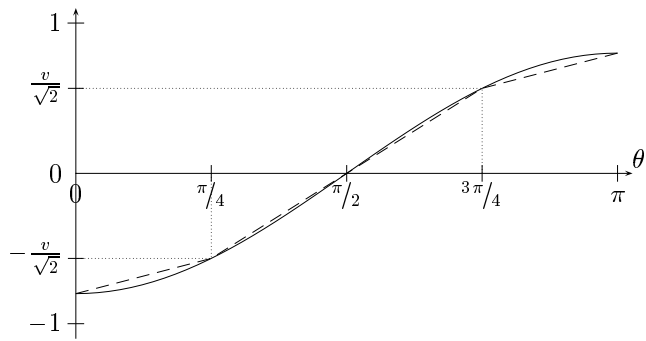}
    \caption{The points where the correlation is tested in the CHSH 
      inequality. Note the small difference between the sinusoidal
      curve and the dashed curve consisting of line segments.}
    \label{fig:point}
  \end{center} 
\end{figure}

In Fig.~\ref{fig:nv1}, there is a region where the local-variable
model does not work which is not ruled out by the CHSH inequality. A
reason for this may be that the correlation is not tested as a
function in the CHSH inequality, but only the values at certain points
are used (see Fig.~\ref{fig:point}).

\begin{figure}[htbp]
  \begin{center}
    \psfig{file=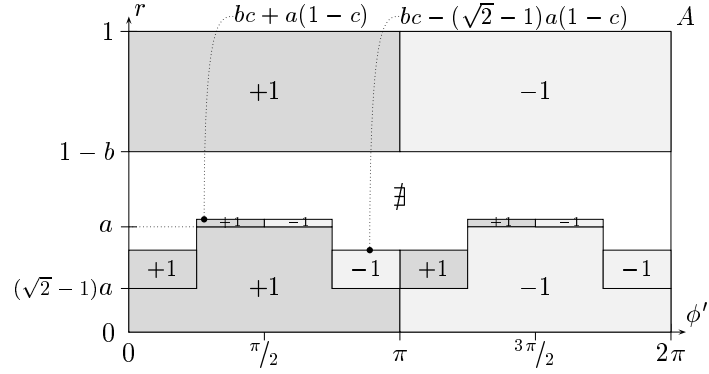} 
    \caption{The detector pattern for detector 1 in the straight-line 
      model with lowered visibility. The detector pattern for detector
      2 is chosen symmetrically.}
    \label{fig:line1}
  \end{center} 
\end{figure}

It would be interesting to know if the straight-line correlation (the
dashed line in Fig.~\ref{fig:point}) is possible to model in the empty
region in Fig.~\ref{fig:nv1}. To construct a model for this case, one
uses the procedure outlined in the previous sections, where the
derivative of the correlation is used to determine the general outline
of the pattern. The correlation is composed of straight lines, so
therefore the pattern should be staircased (i.e., constant
derivative). The derivative should be low from $0$ to
$\textfrac{\pi}4$, then somewhat higher from $\textfrac{\pi}4$ to
$\textfrac{3\pi}4$, and then again low from $\textfrac{3\pi}4$ to
$\pi$. The relation between the derivatives is
\begin{equation}
  \frac{1-\frac1{\sqrt2}}{\frac{\pi}4}\Bigg/
  \frac{\frac1{\sqrt2}}{\frac{\pi}4}=\sqrt2-1.
\end{equation}
The procedure continues, and eventually yields (after addition of
random errors to lower the visibility) the construction in
Fig.~\ref{fig:line1}.

Here, our equations are
\begin{gather}
  \eta=b+\bigl(bc+\frac{a(1-c)}{\sqrt2}\bigr),\\
  \eta^2=2\bigl(bc+\frac{a(1-c)}{\sqrt2}\bigr),
\end{gather}
and
\begin{equation}
  v=\sqrt2 a\Big/\eta^2,
\end{equation}
and we arrive at
\begin{equation}
  \left\{
  \begin{split}
    a&=\frac{v\eta^2}{\sqrt2}\;\; 
    \Bigl(=\frac{2\sqrt2v\eta^2}{4}\Bigr)\\
    b&=\eta-\frac{\eta^2}{2}\\
    c&=\frac{\eta(1-v)}{2-\eta(1+v)}.
  \end{split}
  \right.
\end{equation}
Note that the only change from eqn.~(\ref{eq:abc}) is that the
constant $\pi$ in the expression for $a$ is changed to $2\sqrt2$. This
yields the allowed range as
\begin{equation}
  v\le \frac{1}{2\sqrt2}\bigl(\frac{4}{\eta}-2\bigr).
\end{equation}
And this is exactly the bound from the CHSH inequality.

\begin{figure}[htbp]
  \begin{center}
    \psfig{file=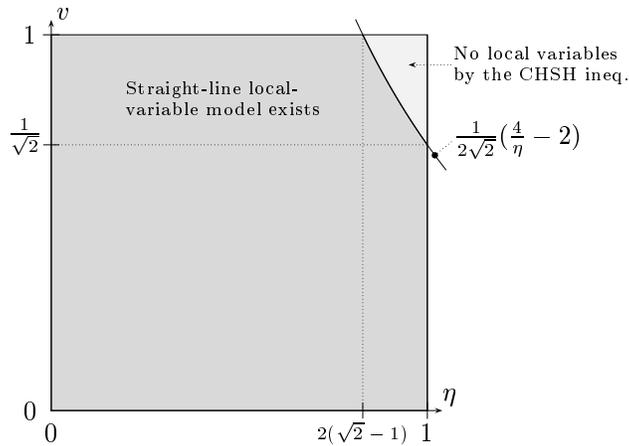} 
    \caption{The region where the straight-line model is
      usable, as compared to the region where no local hidden-variable
      model is possible.}
    \label{fig:nv2}
  \end{center} 
\end{figure}

In essence, we have an explicit model proving the necessity of
having $\eta> 82.83\%$ to contradict local realism in
the CHSH inequality. The bound was shown to be necessary and
sufficient already in Ref.~\cite{GargMerm}, but here an explicit
counter-example is obtained below the bound. In addition, the effect
of lowered visibility is included in this construction.

The same construction for the straight-line case allows a higher
visibility at a given efficiency than for the sinusoidal case. It
seems natural that the sinusoidal form of the correlation is somewhat
more difficult to model than the straight-line form. This enables us
to conjecture that the bound would be lowered if a result that
utilizes the full sinusoidal form of the correlation would be used
instead of the CHSH inequality.

\section{Conclusions} \label{sec:concl}

We have a local-variable model yielding the sinusoidal probabilities
obtained from quantum mechanics, valid for visibilities in the range
\begin{equation}
  v\le \frac{1}{\pi}\big(\frac{4}{\eta}-2\big).\tag{\ref{eq:vis}}
\end{equation}

Experimental results to date may be described using this model by
using the proper values of efficiency and visibility obtained in the
experiment. Although the visibility is high in present experiments,
the efficiency is still too low to yield a decisive test whether
Nature violates local realism or not (if preferred, whether Nature
violates the Bell inequality or not), e.g., in \cite{WJSWZ} the
visibility $v\approx97\%$ and the efficiency $\eta\approx5\%$.

The model is well-behaved in that the efficiency does not vary and the
errors are independent, and neither the non-detection errors nor the
wrong-result errors depend on the detector orientation. Thus, there is
no test on the properties of the errors that would invalidate local
variables for the singlet state below the bound (\ref{eq:vis}).

In all, we have a model which has:
\begin{center}
  QM statistics\\
  Local variables\\
%  A simple structure\\
  Independent errors\\
  $78.80\%$ efficiency\\
  $100\%$ visibility\\
\end{center}
or, if preferred:
\begin{center}
  QM statistics\\
  Local variables\\
%  A simple structure\\
  Independent errors\\
  $63.66\%$ visibility\\
  $100\%$ efficiency\\
\end{center}
and there is currently no experiment result that invalidates this
model.

The question remains if there is a possibility to devise a model where
the bound (\ref{eq:vis}) is on the level of the CHSH inequality, while
keeping the sinusoidal form of the correlation.  By the result in the
previous section, it is more probable that a result utilizing the
whole sinusoidal form of the correlation function rather than the
values at a few points, would lower the bound to (\ref{eq:vis}). The
efficiency bound would then be lowered to $\eta>78.80\%$ for a
violation of local realism.

\section*{Acknowledgements}
The author would like to thank M.\ \.Zukowski and A.\ Zeilinger for
their support.

%\nocite{*} 
%\bibliography{jalar} \bibliographystyle{unsrt}

\end{document}